\documentclass[10pt,twocolumn,english,showpacs,showkeys,superscriptaddress,citeautoscript]{revtex4-2}
\usepackage[T1]{fontenc}
\usepackage{textcomp}
\usepackage[latin9]{inputenc}
\usepackage[active]{srcltx}
\usepackage{babel}
\usepackage{amsmath}
\usepackage{amssymb}
\usepackage{graphicx}
\usepackage[a4paper]{geometry}
\usepackage[pdfusetitle,
 bookmarks=true,bookmarksnumbered=false,bookmarksopen=false,
 breaklinks=false,pdfborder={0 0 1},backref=page,colorlinks=false]
 {hyperref}

\makeatletter

\newcommand{\lyxmathsym}[1]{\ifmmode\begingroup\def\b@ld{bold}
  \text{\ifx\math@version\b@ld\bfseries\fi#1}\endgroup\else#1\fi}

\usepackage{babel}

\makeatother

\begin{document}
\title{Thermodynamic constraints and pseudotransition behavior in a one-dimensional
water-like system}
\author{F. F. Braz}
\address{Department of Physics, Institute of Natural Science, Federal University
of Lavras, 37200-900 Lavras-MG, Brazil}
\author{S. M. de Souza}
\address{Department of Physics, Institute of Natural Science, Federal University
of Lavras, 37200-900 Lavras-MG, Brazil}
\author{M. L. Lyra}
\address{Instituto de Fí­sica, Universidade Federal de Alagoas, 57072-970 Maceió,
Alagoas, Brazil}
\author{Onofre Rojas}
\address{Department of Physics, Institute of Natural Science, Federal University
of Lavras, 37200-900 Lavras-MG, Brazil}
\begin{abstract}
We investigate a one-dimensional water-like lattice model with van
der Waals and hydrogen-bond interactions, allowing for particle number
fluctuations through a chemical potential. The model, defined on a
chain with periodic boundary conditions, exhibits three ground-state
phases: gas, bonded liquid, and dense liquid, separated by sharp phase
boundaries in the chemical potential and temperature plane. Using
the transfer matrix method, we derive exact analytical results within
the grand-canonical ensemble and examine the finite-temperature behavior.
The system exhibits clear pseudotransition features, including sharp
but analytic changes in entropy, density, and internal energy, along
with finite peaks in specific heat and correlation length. To assess
the role of thermodynamic constraints, we consider the behavior under
fixed density through a Legendre transformation. This constrained
analysis reveals smoother anomalies, such as entropy kinks and finite
jumps in specific heat, contrasting with the sharper grand-canonical
signatures. These results underscore the ensemble dependence of pseudotransitions
and show how statistical constraints modulate critical-like behavior.
We also verify that the residual entropy continuity criterion holds
in the grand-canonical ensemble but is violated when the system is
constrained. Our findings illustrate how even a simple one-dimensional
model can mimic water-like thermodynamic anomalies. 
\end{abstract}
\maketitle

\section{Introduction}

Recent studies of decorated one-dimensional spin systems have revealed
sharp thermodynamic anomalies that closely mimic classical phase transitions,
despite the absence of true criticality. Termed pseudotransitions
\citep{pseudo} or ultranarrow crossovers \citep{W-Yin-1,W-Yin-2},
these phenomena arise in a variety of models, including the Ising
diamond chain \citep{W-Yin-2,Strecka-ising,W-Yin-3}, sawtooth-like
geometries \citep{W-Yin-3}, and ladder or triangular tube structures
\citep{W-Yin-1,chapman,hutak21,strk-cav}. Competing interactions,
such as Ising-Heisenberg couplings \citep{torrico,torrico2} and spin-electron
hybrid configurations \citep{Galisova}, often give rise to abrupt
but analytic changes in entropy, density, or specific heat. Similar
signatures have been observed in the extended Hubbard model \citep{psd-hub-dmd},
Potts and Zimm-Bragg-Potts chains \citep{Potts-psd}, and even in
diluted Ising-type systems without explicit decorations \citep{jozef24,Yasinskaya}.
These anomalies occur in systems where different configurations, typically
nearly degenerate or entropically favored, compete within a restricted
phase space. Although true phase transitions are forbidden in one-dimensional
systems with short-range interactions, as established by van Hove
and by Cuesta and Sanchez \citep{Van-hove,Cuesta04}, pseudotransitions
emerge as coherent, nonsingular thermodynamic responses. Beyond their
theoretical significance, such behaviors are relevant for understanding
collective phenomena in real physical systems, including confined
fluids such as water in nanotubes and low-dimensional magnetic or
soft-matter chains.

Several studies \citep{hutak21,unv-cr-exp,krokhmalskii21} have reported
power-law scaling in thermodynamic quantities near pseudo-critical
points, with critical exponents that satisfy the Rushbrooke inequality.
This recurring behavior, which spans classical and quantum models,
underscores geometric frustration and competing energy scales as universal
drivers of pseudotransition thermodynamics. Pseudotransitions have
been further analyzed in reference \citep{Isaac}, focusing on spin
correlation functions. Further investigation of correlation length
was also performed by Chapman-Tomasell-Carr\citep{chapman} for a
toblerone-type lattice. More recently Yasinskaya and Panov\citep{Yasinskaya}
analyzed canonical pseudotransitions in diluted spin chains via Maxwell
construction, and the result contrasts with grand-canonical behavior.
It may also have potential applications in the theoretical understanding
of quantum many-body machines\citep{machines25}.

On the other hand, water's ubiquity in sustaining life, from enabling
solvent interactions in cells to regulating Earth's climate, underscores
its unique role as a matrix for life \citep{Franks2000}. Its anomalous
behavior, such as the density maximum at 4°C and enhanced diffusion
under pressure, has long intrigued scientists \citep{Debenedetti2003,Malenkov2009}.
These deviations from typical liquid behavior are crucial for biological
and environmental processes, motivating efforts to unravel their microscopic
origins. While water's hydrogen-bonding network is central to its
anomalies, simplified models like core-softened potentials and lattice
models have revealed analogous behaviors in diverse fluids, suggesting
universal mechanisms \citep{Errington2001,Oliveira2006a}.

Core-softened potentials model competing interactions in molecular
simulations, revealing density/diffusion anomalies \citep{Barbosa2013}.
While simplified, lattice systems also reproduce these anomalies through
excluded volume and interaction topology \citep{Barbosa2011}. Research
reveals multiple anomalous zones in phase diagrams, suggesting complex
structural shifts \citep{Barbosa2013,Rizzatti2018}. For example,
Barbosa et al. identified two anomaly regions in a 1D core-softened
fluid tied to competing local structures and liquid-liquid transitions
\citep{Barbosa2013}. Similarly, lattice models show anomalies driven
by entropy, with residual entropy linked to density extremes in repulsive
gases \citep{daSilva2015}. Ferreira et al. \citep{caparica19} recently
analyzed a 1D repulsive lattice gas using transfer matrix methods,
exploring dimer density, vacancies, and entropy-driven water-like
anomalies. Another one-dimensional model for water and aqueous solutions
were discussed by Ben-Naim\citep{Arieh}.

This work is organized as follow. In Sect. 2 is discussed the grand-canonical
ensemble (GCE) for 1D water-like model, where we explore rigorously
and analyze the zero-temperature phase diagram and its thermal anomalous
behavior. In Sec. 3, we explore the thermodynamic response at fixed
density, using a Legendre transformation to access quantities as functions
of density and temperature. Finally, in Sec. 4, we present our conclusions
and perspectives.

\section{grand-canonical ensemble of 1D water-like system}

In this one-dimensional lattice model, each site can either be occupied
by a molecule or remain vacant \citep{Barbosa2011}. Two types of
intermolecular interactions are considered, each reflecting distinct
physical mechanisms. The first is a short-range van der Waals attraction,
which acts between nearest-neighbor molecules. This reflects the non-directional,
distance-dependent nature of dispersion forces, which are strongest
when molecules are in immediate contact. The second interaction models
hydrogen bonding, which occurs between next-nearest-neighbor molecules
that are separated by an empty site. This spatial configuration captures
the directional and longer-range character of hydrogen bonds, which
typically require specific geometrical arrangements and can span slightly
greater distances than van der Waals forces. In this model, the presence
of a vacancy between interacting molecules mimics the angular constraints
and excluded volume effects often necessary for hydrogen bond formation.
The effective Hamiltonian in the grand canonical ensemble (GCE) is
given by
\begin{equation}
H\!=\!-\!\sum_{i=1}^{N}\!\left[\epsilon_{v}\eta_{i}\eta_{i+1}+\epsilon_{h}\eta_{i}\!\left(\!1\!-\!\eta_{i+1}\!\right)\!\eta_{i+2}\!+\!\mu\eta_{i}\right],\label{eq:Ham1}
\end{equation}
where $\eta_{j}$ indicates site $j$'s occupation ($\eta_{j}=1$
if occupied, $\eta_{j}=0$ if vacant) of a molecule, $\epsilon_{v}$
is the van der Waals attraction between nearest neighbors, $\epsilon_{h}>0$
is the attractive hydrogen bond energy between next-nearest neighbors
separated by a vacancy, and $\mu$ is the chemical potential. The
positive sign of $\epsilon_{h}$ reflects the typical stabilizing
role of hydrogen bonds in real molecular systems. This coarse-grained
representation abstracts the essential spatial and energetic features
of real molecular interactions, enabling analytical or numerical treatment
while preserving key physical behavior.

In the following, we analyze the thermodynamic properties within the
GCE framework.

\subsection{Transfer matrix and grand potential}

In the GCE, where $T$ and $\mu$ are the natural variables, the thermodynamic
properties can be computed using the transfer matrix method. Due to
the next-nearest-neighbor hydrogen bonding term in the Hamiltonian
\eqref{eq:Ham1}, the local state of a site depends on two neighboring
sites. This motivates a four-state basis constructed from site pairs:
(00), (01), (10), and (11), where each digit indicates vacancy or
occupation.

In this basis, the transfer matrix takes the form:

\begin{equation}
\mathbf{V}=\left(\begin{array}{cccc}
1 & 1 & 0 & 0\\
0 & 0 & 1 & 1\\
z & bz & 0 & 0\\
0 & 0 & az & az
\end{array}\right),\label{eq:TM1}
\end{equation}
where $a={\rm e}^{\beta\epsilon_{v}}$, $b={\rm e}^{\beta\epsilon_{h}}$,
$z={\rm e}^{\beta\mu}$, and $\beta=\frac{1}{k_{B}T}$, with $T$
being the absolute temperature and $k_{B}$ the Boltzmann constant.

The eigenvalues of this matrix determine the thermodynamic behavior
in the thermodynamic limit. The non-trivial characteristic equation
is a cubic

\begin{equation}
\lambda^{3}-(az+1)\lambda^{2}-z(a-b)\lambda-z(b-1)=0.\label{eq:sec-eq}
\end{equation}
The three roots of this cubic equation can be conveniently written
via trigonometric functions:

\begin{equation}
\lambda_{j}=2\sqrt{Q}\cos\left(\tfrac{\theta-2\pi j}{3}\right)+\tfrac{az+1}{3},\label{eq:L_j}
\end{equation}
for $j=0,1,2$, where

\begin{alignat}{1}
\theta= & \arccos\Bigl(\tfrac{R}{\sqrt{Q^{3}}}\Bigr),\\
Q= & \left(\tfrac{az+1}{3}\right)^{2}+\tfrac{z}{3}\left(b-a\right),\nonumber \\
= & \tfrac{a^{3}z^{3}+1}{9\left(az+1\right)}+\tfrac{bz}{3}>\tfrac{1}{9},\\
R= & \left(\tfrac{az+1}{3}\right)^{3}+\tfrac{z}{2}\left(\tfrac{az+1}{3}\right)\left(b-a\right)+\tfrac{z}{2}\left(1-b\right).
\end{alignat}

Unlike symmetric transfer matrices whose eigenvalues are guaranteed
to be real, the non-symmetric matrix in Eq \eqref{eq:L_j} can yield
complex eigenvalues, in contrast with many well-studied models \citep{pseudo,W-Yin-1,Strecka-ising,jozef24,Yasinskaya,on-strk,hub-1d},
where the spectrum is typically real. Since $Q>1/9$ is always positive
while R can take any real value, two scenarios arise: if $R^{2}\leqslant Q^{3}$,
all eigenvalues are real; if $R^{2}>Q^{3}$, the equation yields one
real root and a pair of complex conjugates. This behavior is confirmed
across different parameter values.

We first analyze the case $R^{2}<Q^{3}$, where all three eigenvalues
$\lambda_{j}$ are real and distinct.

For a system with $N$ sites, the grand partition function takes the
form

\begin{equation}
\Xi_{N}=\lambda_{0}^{N}\left(1+\frac{\lambda_{1}^{N}}{\lambda_{0}^{N}}+\frac{\lambda_{2}^{N}}{\lambda_{0}^{N}}\right).
\end{equation}
Since $\lambda_{0}>\lambda_{1}>\lambda_{2}$, the correction terms
vanish exponentially as $N\to\infty$, and the grand potential per
site reduces to

\begin{alignat}{1}
\omega= & -\frac{1}{\beta}\ln(\lambda_{0}).\label{eq:Phi}
\end{alignat}
In the special case $R^{2}=Q^{3}$, where $\theta=0$, the eigenvalues
reduce to

\begin{alignat}{1}
\lambda_{0}= & 2\sqrt{Q}+\tfrac{az+1}{3}\\
\lambda_{1}=\lambda_{2}= & -\sqrt{Q}+\tfrac{az+1}{3},
\end{alignat}
The grand partition function for $N$ sites, becomes

\begin{equation}
\Xi_{N}=\lambda_{0}^{N}\left(1+2\frac{\lambda_{1}{}^{N}}{\lambda_{0}^{N}}\right),
\end{equation}
which leads to the same expression for the grand potential in the
thermodynamic limit. In both cases, the leading eigenvalue $\lambda_{0}$
alone determines the bulk thermodynamics.

In the parameter regimes of interest, all eigenvalues are real. Therefore,
we do not further explore the complex root scenarios, as they lie
beyond the scope of this study.

From the grand potential, several thermodynamic quantities of interest
can be directly obtained: the entropy per site $\mathcal{S}=-\partial\omega/\partial T$,
the molecular density $\rho=-\partial\omega/\partial\mu$, the specific
heat per site $C=T\partial\mathcal{S}/\partial T$, and the correlation
length $\xi=1/\ln(\lambda_{0}/\lambda_{1})$.

\subsection{Phase diagram and residual entropy}

According to previous studies, this model is expected to exhibit three
distinct phases at zero temperature: a bonded fluid (BF), a dense
fluid (DF), and a gas (G) phase\citep{Barbosa2013}.

\begin{figure}[h]
\includegraphics[scale=0.45]{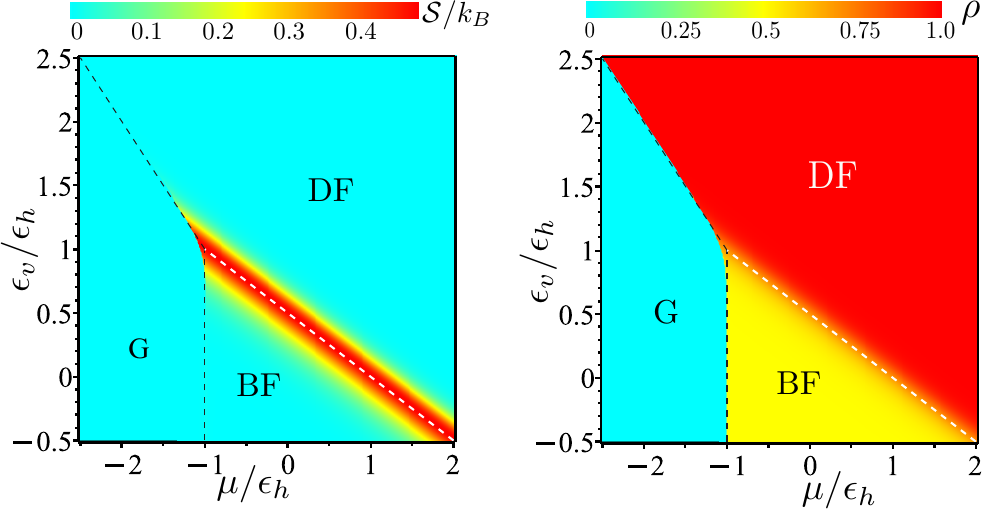}\caption{Phase diagram $\epsilon_{v}/\epsilon_{h}$ against $\mu/\epsilon_{h}$.
(Left) The background density plot corresponds to the entropy ${\cal S}/k_{B}$
for a low temperature value $k_{B}T/\epsilon_{h}=0.1$. (Right) The
density $\rho$ plot describes the molecule density at $k_{B}T/\epsilon_{h}=0.1$.}
\label{fig:Phase-diagram} 
\end{figure}

In Fig.\,\ref{fig:Phase-diagram}, we present the zero-temperature
phase diagram in the $\mu/\epsilon_{h}\lyxmathsym{\textendash}\epsilon_{v}/\epsilon_{h}$
plane, denoted by dashed line. It is worth noting that there is no
residual boundary entropy along the G--DF and G--BF interfaces.
However, the residual entropy at the BF--DF boundary is given by
\begin{equation}
\mathcal{S}/k_{B}=\ln\left(\tfrac{\sqrt{5}+1}{2}\right)=0.481211825,
\end{equation}
which can be obtained either by taking the limit $T\rightarrow0$
or by counting the number of accessible microstates (see Section \ref{sec:3}).
This behavior is clearly observed in the background of the left panel,
which illustrates the entropy at low temperature, specifically for
$k_{B}T/\epsilon_{h}=0.1$. The right panel shows the thermal density
of molecules as a background at the same temperature, where we identify
regions with $\rho=0$ (cyan), $\rho=0.5$ (yellow), and $\rho=1$
(red). Notably, the boundary between the G-BF and G-DF phases is sharp,
due to the absence of residual entropy. In contrast, the BF-DF boundary
appears smoother, as residual entropy is present in this region.

\subsection{Pseudo-critical temperature}

From Fig.\ref{fig:Phase-diagram}, we observe anomalous behavior around
the point where the three phases meet. To examine this more rigorously,
we consider the system near this region, where thermodynamic observables
such as the specific heat, or correlation length display sharp but
finite peaks at a temperature known as the pseudo-transition point
$T_{p}$. Although these features resemble second-order phase transitions,
all thermodynamic functions remain analytic. In our model, the pseudo-criticality
arises from a near-degeneracy of dominant eigenvalues of the transfer
matrix, producing quasi-singular behavior in $C$ and $\xi$ without
true nonanalyticity in the free energy.

First, let us analyze the eigenvalues of the transfer matrix in a
particular case: When $\mu<0$, we have $z\ll1$ (in the limit of
low temperature). Under this condition the transfer matrix \eqref{eq:TM1}
simplifies to 
\begin{equation}
V=\left(\begin{array}{cccc}
1 & 1 & 0 & 0\\
0 & 0 & 1 & 1\\
0 & bz & 0 & 0\\
0 & 0 & az & az
\end{array}\right).
\end{equation}
In this limit, the characteristic polynomial ${\rm det}\left(\mathbf{V}-\lambda\right)$
reduces to 
\begin{equation}
\lambda\left(\lambda-1\right)\left(\lambda^{2}-za\lambda-bz\right)=0,\label{eq:seq-eq-spd}
\end{equation}
and the corresponding eigenvalues are 
\begin{equation}
\lambda=\left\{ 0,1,\tfrac{az}{2}\pm\tfrac{\sqrt{a^{2}z^{2}+4bz}}{2}\right\} .
\end{equation}
It is interesting to note that, in this limit, the largest eigenvalue
can be expressed as 
\begin{equation}
\lambda={\rm max}\left(1,\tfrac{az}{2}+\tfrac{\sqrt{a^{2}z^{2}+4bz}}{2}\right),
\end{equation}
where the maximum arises from competing eigenvalues, similar to what
was observed in reference \citep{pseudo}. In this limiting case,
we are effectively forcing the emergence of a ``phase transition''
$z\rightarrow0$. This leads to a critical-like condition given by
\begin{equation}
1=\frac{az}{2}+\frac{\sqrt{a^{2}z^{2}+4bz}}{2},
\end{equation}
which, when solved, yields the simple relation $z^{-1}=a+b$. This
can be further rewritten using a transcendental equation that defines
the pseudo-critical temperature $T_{p}$: 
\begin{equation}
T_{p}=\frac{-(\mu+\epsilon_{h})/k_{B}}{\ln\left(1+{\rm e}^{\frac{\epsilon_{v}-\epsilon_{h}}{k_{B}T_{p}}}\right)}.\label{eq:Tp-mu}
\end{equation}
This condition identifies a characteristic temperature $T_{p}$ ,
referred to as the pseudo-critical temperature \citep{pseudo}, since
a true phase transition does not occur in this model when the full
solution of Eq.\eqref{eq:L_j} is considered rigorously.

From the above result, one can generalize the condition for the appearance
of a pseudotransition in terms of the eigenvalues: when $\lambda_{0,p}+\lambda_{1,p}=2$,
as explored in Refs. \citep{hub-1d,pimenta-22,Bjp-rojas20}, the condition
becomes

\begin{alignat}{1}
3\sqrt{Q_{p}}\cos\left(\tfrac{\theta_{p}-\pi}{3}\right)= & (a_{p}z_{p}+4),
\end{alignat}
which represents a more general criterion for this model. However,
in practice, it is sufficient to consider the condition given in \eqref{eq:Tp-mu}.

\begin{figure}
\includegraphics[scale=0.56]{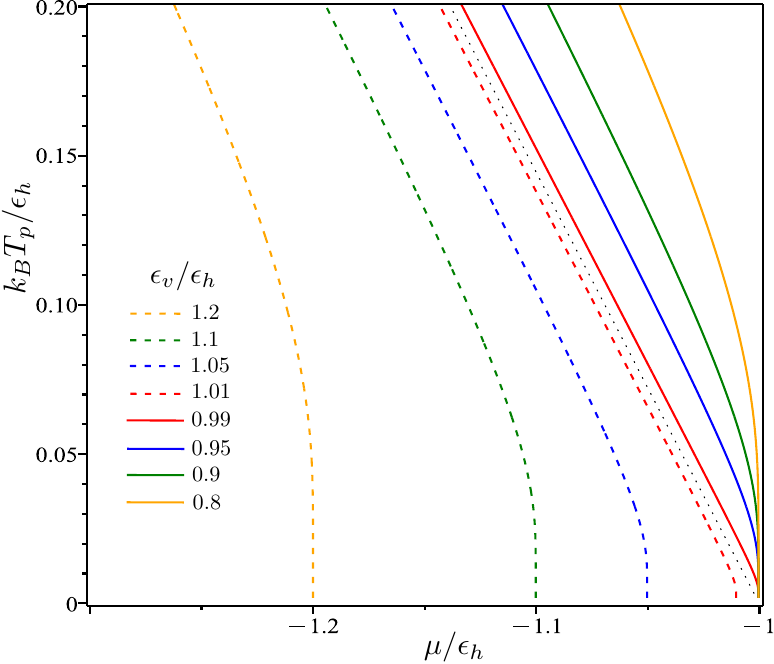}\caption{Pseudo-critical temperature $k_{B}T_{p}/\epsilon_{h}$ as a function
of the chemical potential $\mu/\epsilon_{h}$, calculated using the
expression given in \eqref{eq:Tp-mu}.}
\label{fig:Pseudo-critical-T} 
\end{figure}

Figure \ref{fig:Pseudo-critical-T} shows the pseudo-critical temperature
$k_{B}T_{p}/\epsilon_{h}$ as a function of the chemical potential
$\mu/\epsilon_{h}$ for several values of $\epsilon_{v}/\epsilon_{h}$,
indicated in the legend. The solid lines correspond to $\epsilon_{v}/\epsilon_{h}<1$,
where all curves converge to the same chemical potential value as
$T_{p}\to0$. The dashed lines represent $\epsilon_{v}/\epsilon_{h}>1$;
in this regime, $T_{p}$ also tends to zero, but the corresponding
chemical potential becomes more negative ($\mu/\epsilon_{h}<-1$).
The dotted line corresponds to the particular case $\epsilon_{v}/\epsilon_{h}=1$,
for which both $T_{p}\to0$ and $\mu/\epsilon_{h}\to-1$.

\begin{figure}
\includegraphics[scale=0.6]{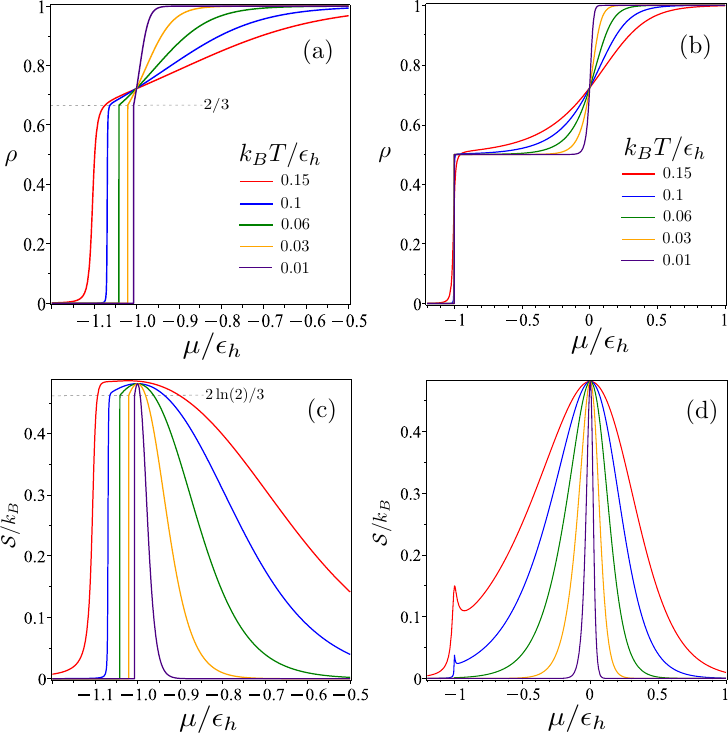}\caption{(a) Density as a function of chemical potential, assuming $\epsilon_{v}/\epsilon_{h}=1$.
(b) Density as a function of chemical potential, for $\epsilon_{v}/\epsilon_{h}=0.5$
. (c) Entropy as a function of chemical potential, for fixed $\epsilon_{v}/\epsilon_{h}=1$.
(d) Entropy as a function of chemical potential, considering $\epsilon_{v}/\epsilon_{h}=0.5$. }
\label{fig:rho-S-mu} 
\end{figure}

In Fig.\,\ref{fig:rho-S-mu}a, the density $\rho$ is shown as a
function of $\mu/\epsilon_{h}$ for several fixed temperatures (indicated
in the legend), assuming $\epsilon_{v}/\epsilon_{h}=1$. All curves
intersect at the same point, $\mu/\epsilon_{h}=-1$, where the density
is

\begin{equation}
\rho=\tfrac{1}{2}+\tfrac{\sqrt{5}}{10}=0.7236068,
\end{equation}
independent of temperature. This intersection is clearly illustrated
in Fig.\,\ref{fig:rho-S-mu}a. At zero temperature, this special
point corresponds to the meeting of all three phases. Additionally,
another anomaly appears as a kink in the density curves, consistently
occurring near $\rho=2/3$ in the low-temperature regime. This density
was previously identified in the literature \citep{Barbosa2011} and
appears almost independently of temperature, although the kink becomes
less pronounced at higher temperatures. Notably, in this region, the
density curve is nearly a vertical straight line up to $\rho=2/3$,
where a kink emerges precisely at the pseudo-critical temperature
$T_{p}$. This indicates that the pseudo-critical temperature occurs
at an almost constant density of $\rho=2/3$. The above characteristic
densities will be explicitly discussed in the next section.

In Fig. \ref{fig:rho-S-mu}b, we present an analysis similar to panel
(a), but for a different interaction ratio, $\epsilon_{v}/\epsilon_{h}=0.5$,
which corresponds to the boundary between the bonded fluid (BF) and
dense fluid (DF) phases. At $\mu=0$, the density is observed to be
temperature-independent, occurring precisely at the boundary between
the quasi-gas (qG) and quasi-dense fluid (qDF) regions (defined as
in \citep{pseudo}), at $\rho=0.7236068$. For $\mu\gtrsim0.8$, the
density $\rho$ remains nearly constant across temperatures.

Under the same conditions as in panel (a), Fig. \ref{fig:rho-S-mu}c
shows the entropy as a function of chemical potential $\mu/\epsilon_{h}$,
where we can observe anomalous behavior. First, the entropy reaches
a maximum nearly at $\mu/\epsilon_{h}\approx-1$, with a constant
magnitude of $S/k_{B}=0.481211825k_{B}$ as $k_{B}T/\epsilon_{h}\rightarrow0$,
strongly correlated with the density $\rho=0.7236068$. At higher
temperatures, this maximum slightly shifts. The second anomaly appears
as a kink in the entropy curve at ${\cal S}/k_{B}=2\ln(2)/3\approx0.462098$,
associated with $\rho=2/3$. These characteristic entropies will be
further addressed in section \ref{sec:3}. Just below this kink, the
entropy curve becomes almost a vertical straight line, indicating
the pseudotransition point.

To complete our analysis, Fig. \ref{fig:rho-S-mu}d displays the entropy
under the same conditions as in panel (b), as a function of $\mu/\epsilon_{h}$.
A residual entropy of $S/k_{B}=0.481211825$ is observed at $\mu/\epsilon_{h}=0$,
marking the crossover between the quasi-bonded fluid (qBF) and qDF
regions. Meanwhile, at $\mu/\epsilon_{h}=-1$, which corresponds to
the boundary between the qG and qBF regions (defined as in \citep{pseudo}),
there is no residual entropy at zero temperature.

\begin{figure}
\includegraphics[scale=0.6]{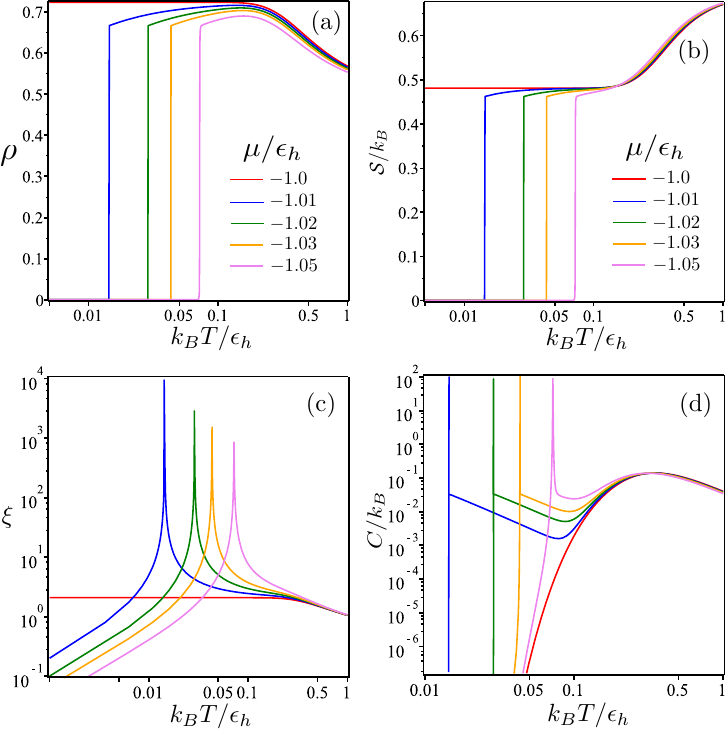}\caption{(a) Density as a function of temperature, assuming $\epsilon_{v}/\epsilon_{h}=1$.
(b) Entropy as a function of temperature, for the same set of parameters
and conditions as in panel (a). (c) Correlation length $\xi$ as a
function of temperature, under the same conditions as in panel (a).
(d) Specific heat $C$ as a function of temperature, using the same
set of parameters as in the previous panels.}
\label{fig:Qnts-T} 
\end{figure}

In Fig.\ref{fig:Qnts-T}, panel (a) shows the density $\rho$ as a
function of temperature, assuming $\epsilon_{v}/\epsilon_{h}=1$,
for several values of chemical potential with $\mu/\epsilon_{h}<-1$.
As in our previous analysis, the density increases sharply, resembling
a vertical line, up to $\rho=2/3$ at $k_{B}T_{p}/\epsilon_{h}$ for
each curve, where we observe a kink. The density then increases further,
reaching a maximum of $\rho=0.7236068$. Note that for $\mu/\epsilon_{h}=-1$,
there is no pseudotransition at finite temperature, as indicated by
the red curve. Panel (b) shows the entropy as a function of temperature.
Here, we observe a typical jump (although the curve remains continuous)
at $T_{p}$ , reaching an entropy of ${\cal S}/k_{B}=2\ln(2)/3\approx0.462098$,
followed by a kink in the curvature, eventually stabilizing at a plateau
value of $\mathcal{S}/k_{B}=0.481211825$. For higher temperatures,
the entropy follows standard behavior. A similar analysis is shown
for the correlation length $\xi$ as a function of temperature in
panel (c), where we observe sharp peaks at the pseudo-critical temperature
$k_{B}T_{p}/\epsilon_{h}$, resembling the behavior seen in a second-order
phase transition, marking the boundary between the qG region and coexistence
with the qDF and qBF regions. Panel (d) illustrates the specific heat
$C$ as a function of temperature. Here, we also observe the characteristic
behavior of a pseudotransition, with a very sharp peak at $T_{p}$.
However, the anomalous behavior observed at $\rho=0.7236068$ is not
clearly manifested in the last two panels (c-d).

\section{Thermodynamic analysis at fixed density }

\label{sec:3}

Let us now consider the system from a constrained perspective, where
temperature $T$ and particle density $\rho$ are treated as the natural
thermodynamic variables. To access this description, we perform a
Legendre transformation of the grand potential to obtain the Helmholtz
free energy per site, $f(T,\rho)$. The transformation reads

\begin{equation}
f(T,\rho)=\max_{\mu}\left[\omega(T,\mu)+\rho\,\mu\right].
\end{equation}
Since there is no true phase transition, the maximization procedure
is unnecessary. To obtain the Helmholtz free energy density $f(T,\rho)$,
we can eliminate the chemical potential $\mu$ by using the thermodynamic
relation $\rho=-\partial\omega/\partial\mu$, which, in terms of the
fugacity $z=e^{\beta\mu}$, can be rewritten as $\frac{\partial z}{\partial\lambda}=\frac{z}{\lambda\rho}$.
In principle, this allows us to express $\mu$ as a function of $\rho$,
i.e., $\mu=\mu(\rho)$. Substituting this into the expression for
the grand potential density yields the Helmholtz free energy

\begin{equation}
f(T,\rho)=\omega(T,\mu(\rho))+\rho\,\mu(\rho).
\end{equation}

Part of this transformation was previously explored in Ref. \citep{Barbosa2011},
but here we will extend the analysis. Theoretically, this transformation
can always be performed; however, the analytical process can be highly
challenging, even for relatively simple models. Most of this transformation
can be performed numerically \citep{Yasinskaya}. Analytical transformations
occur only in rare cases, but we are lucky that this model can be
transformed analytically, although the transformation involves cumbersome
algebraic expressions.

\subsection{Pseudo-critical temperature dependent on $\rho$}

Let us now consider the specific transformation of $z$ as a function
of $\rho$, particularly around the pseudotransition point. In this
simplified case, we start from the expression given in \eqref{eq:seq-eq-spd},
which allows us to write the relation: 
\begin{equation}
z=\frac{\lambda^{2}}{a\lambda+b}.\label{eq:z(lmbd)-psd}
\end{equation}

Following the procedure outlined earlier, we differentiate the quadratic
polynomial of Eq. \eqref{eq:seq-eq-spd} with respect to $\lambda$,
and use the relation 
\begin{equation}
\frac{\partial z}{\partial\lambda}=\frac{z}{\lambda\rho}.
\end{equation}
Substituting this into Eq. \eqref{eq:z(lmbd)-psd}, we obtain the
expression 
\begin{equation}
a\left(\rho-1\right)\lambda^{2}+b\left(2\rho-1\right)\lambda=0.
\end{equation}
This yields two solutions: a trivial one, $\lambda=0$, and a non-trivial
solution given by 
\begin{equation}
\lambda=\frac{b\left(2\rho-1\right)}{a\left(1-\rho\right)},\quad\text{for}\quad\tfrac{a+b}{a+2b}<\rho<1,
\end{equation}
where the lower bound restriction of $\rho$ is obtained in the limit
$\lambda\rightarrow1$.

Using this result, we can explicitly invert and express $z$ in terms
of $\rho$ as 
\begin{equation}
z=\frac{b\left(2\rho-1\right)^{2}}{a^{2}\left(1-\rho\right)\rho}.
\end{equation}
Therefore, the Helmholtz free energy can be written explicitly as
\begin{alignat}{1}
f(T,\rho)= & -k_{B}T\ln\left(\tfrac{b\left(2\rho-1\right)}{a\left(1-\rho\right)}\right)+\rho k_{B}T\ln(z)\\
= & -(1-\rho)\epsilon_{h}-(2\rho-1)\epsilon_{v}+\nonumber \\
 & -k_{B}T\ln\left[\frac{\left(2\rho-1\right)^{\left(1-2\rho\right)}\rho^{\rho}}{\left(1-\rho\right)^{\left(1-\rho\right)}}\right].
\end{alignat}
In this limit, we can apply the condition $z^{-1}=a+b$, previously
derived, which allows us to express the pseudo-critical temperature
$T_{p}$ as a function of $\rho$: 
\begin{equation}
T_{p}=\frac{\epsilon_{v}-\epsilon_{h}}{k_{B}\ln\left(\frac{2\rho-1}{1-\rho}\right)}.\label{eq:Tp-rho}
\end{equation}
This expression is equivalent to the result given in \eqref{eq:Tp-mu}.

\subsection{Helmholtz free energy}

We can apply a similar transformation as before, this time to express
the largest eigenvalue from \eqref{eq:sec-eq} purely in terms of
$\rho$, eliminating the fugacity $z$. Starting from the secular
cubic polynomial given in \eqref{eq:sec-eq}, we assume that $z$
depends on $\lambda$, allowing us to relate it to $\rho$. This relationship
was previously explored in reference \citep{Barbosa2011} and is given
by: 
\begin{equation}
z=\frac{\lambda^{2}\left(\lambda-1\right)}{a\,\lambda^{2}-a\lambda+b\lambda-b+1}.\label{eq:z(lambda)}
\end{equation}

Taking the derivative of \eqref{eq:sec-eq} with respect to $\lambda$
and using the identity $\frac{\partial z}{\partial\lambda}=\frac{z}{\lambda\rho}$,
we can substitute it into \eqref{eq:z(lambda)} to eliminate the dependence
of $z$. After some algebraic manipulation, this leads to the cubic
polynomial in $\lambda$: 
\begin{alignat}{1}
\mathfrak{a}_{3}\lambda^{3}+\mathfrak{a}_{2}\lambda^{2}+\mathfrak{a}_{1}\lambda+\mathfrak{a}_{0} & =0,\label{eq:pol3-lmbd}
\end{alignat}
with the coefficients defined as 
\begin{alignat}{1}
\mathfrak{a}_{0}= & \left(2\rho-1\right)\left(b-1\right),\\
\mathfrak{a}_{1}= & 2\left(1-2\rho\right)b+a\left(\rho-1\right)+3\rho-1,\\
\mathfrak{a}_{2}= & \left(2\rho-1\right)b+2\left(1-\rho\right)a,\\
\mathfrak{a}_{3}= & \left(\rho-1\right)a.
\end{alignat}

Note that the cubic polynomial equation now depends only on the parameters
$a$, $b$, and $\rho$, and that the chemical potential $\mu$ has
been eliminated. Although Eq. \eqref{eq:pol3-lmbd} is not equivalent
to the original cubic polynomial \eqref{eq:sec-eq}, only their largest
roots are equivalent, since $\rho$ is defined using the largest root
of \eqref{eq:sec-eq}.

Thus, the largest root of Eq. \eqref{eq:pol3-lmbd} can be written
as: 
\begin{equation}
\lambda_{0}=2\sqrt{\tilde{Q}}\cos\left(\tfrac{\tilde{\phi}}{3}\right)+\tfrac{\mathfrak{a}_{2}}{3},\label{eq:L_0}
\end{equation}
where 
\begin{alignat}{1}
\tilde{\phi}= & \arccos\left(\tfrac{\tilde{R}}{\sqrt{\tilde{Q}^{3}}}\right)\\
\tilde{Q}= & \left(\tfrac{\mathfrak{a}_{2}}{3}\right)^{2}-\tfrac{\mathfrak{a}_{1}}{3},\\
\tilde{R}= & \tfrac{\mathfrak{a}_{1}\mathfrak{a}_{2}}{6}-\tfrac{\mathfrak{a}_{0}}{2}-\left(\tfrac{\mathfrak{a}_{2}}{3}\right)^{3},
\end{alignat}

Furthermore, by combining \eqref{eq:sec-eq} and \eqref{eq:pol3-lmbd},
we can eliminate $\lambda$ and write the fugacity $z$ purely as
a function of $\rho$. This yields a cubic equation in $z$: 
\begin{alignat}{1}
z^{3}+\mathfrak{b}_{2}z^{2}+\mathfrak{b}_{1}z+\mathfrak{b}_{0} & =0,\label{eq:pol3-lmbd-1}
\end{alignat}
where the coefficients are given by 
\begin{alignat}{1}
\mathfrak{b}_{0}= & \frac{\left(2\rho-1\right)^{2}\left(b-1\right)}{\rho a^{2}\mathfrak{c}_{3}\left(\rho-1\right)},\label{eq:bfk0}\\
\mathfrak{b}_{1}= & \frac{\mathfrak{c}_{1}}{a^{2}\mathfrak{c}_{3}}-\frac{\left(b-1\right)\left(2\rho\left(a+b\right)-9\rho+1\right)}{\rho^{2}a^{2}\mathfrak{c}_{3}\left(\rho-1\right)},\\
\mathfrak{b}_{2}= & -\frac{\mathfrak{c}_{2}}{a^{2}\mathfrak{c}_{3}}-\frac{b}{\rho a^{2}\left(\rho-1\right)},\label{eq:bfk2}
\end{alignat}
with $\mathfrak{c}_{i}$ depending on $a$ and $b$, defined as 
\begin{alignat}{1}
\mathfrak{c}_{1}= & a^{2}-8ab-8b^{2}+6a+36b-27,\\
\mathfrak{c}_{2}= & 2a^{3}-2b\,a^{2}-8a\,b^{2}-4b^{3}-6a^{2}+18ab,\\
\mathfrak{c}_{3}= & \left(a+b\right)^{2}-4a.
\end{alignat}
We can now solve this cubic equation using the same method as outlined
in the preceding section. Although the full solution is algebraically
tractable, it results in a cumbersome expression that will not be
displayed explicitly here.

The roots of the cubic equation can be formally written as 
\begin{equation}
z_{j}=2\sqrt{\mathfrak{Q}}\cos\left(\tfrac{\vartheta-2\pi j}{3}\right)+\tfrac{\mathfrak{b}_{2}}{3},\label{eq:L_j-1}
\end{equation}
where 
\begin{alignat}{1}
\vartheta= & \arccos\left(\tfrac{\mathfrak{R}}{\sqrt{\mathfrak{Q}^{3}}}\right)\\
\mathfrak{Q}= & \left(\tfrac{\mathfrak{b}_{2}}{3}\right)^{2}-\tfrac{\mathfrak{b}_{1}}{3},\\
\mathfrak{R}= & \tfrac{\mathfrak{b}_{1}\mathfrak{b}_{2}}{6}-\tfrac{\mathfrak{b}_{0}}{2}-\left(\tfrac{\mathfrak{b}_{2}}{3}\right)^{3},
\end{alignat}
with the coefficients $\mathfrak{b}_{i}$ defined in (\ref{eq:bfk0}-\ref{eq:bfk2}).

Although the general solution can be written algebraically, the resulting
expressions in terms of $a$, $b$ and $\rho$ are quite lengthy.
Therefore, we present it in a more compact and formal form. 
\begin{alignat}{1}
f(T,\rho)= & -k_{B}T\ln\left(\lambda_{0}\right)+\rho k_{B}T\ln(z_{0}),
\end{alignat}
where $\lambda_{0}$ is given in \eqref{eq:L_0} and $z_{0}$ is obtained
from \eqref{eq:L_j-1}, both depending only on $a$, $b$ and $\rho$.

From the Helmholtz free energy, one can calculate the entropy as $\mathcal{S}(T,\rho)=-\partial f(T,\rho)/\partial T$,
among other thermodynamic quantities.

\begin{figure}
\includegraphics[scale=0.45]{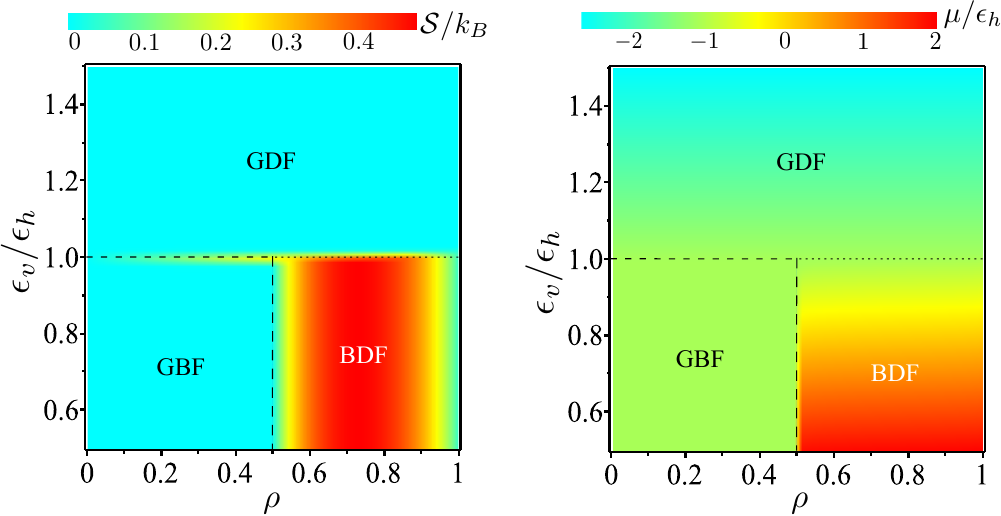}\caption{Phase diagram $\epsilon_{v}/\epsilon_{h}$ versus $\rho$. The dotted
line corresponds to the DF phase, the short-dashed line represents
the G phase, and the long-dashed line denotes the BF phase. (Left)
The background density plot shows the entropy ${\cal S}/k_{B}$ for
$k_{B}T/\epsilon_{h}=0.01$. (Right) The background shows $\mu/\epsilon_{h}$
for the same temperature.}
\label{fig:Phase-diagram-rho} 
\end{figure}

Fig. \ref{fig:Phase-diagram-rho} illustrates the zero-temperature
phase diagram in the $\rho\text{\textendash}\epsilon_{v}/\epsilon_{h}$
plane. The dotted line corresponds to the DF phase, the short-dashed
line represents the G phase, and the long-dashed line depicts the
BF phase. These phases were previously identified in Fig. \ref{fig:Pseudo-critical-T}.
In this version of the phase diagram, we observe the coexistence of
the gas--bonded-fluid (GBF) phase for densities restricted to $0<\rho<0.5$
and $\epsilon_{v}/\epsilon_{h}<1$. Similarly, a bonded-dense-fluid
(BDF) coexistence occurs for $0.5<\rho<1$ and $\epsilon_{v}/\epsilon_{h}<1$.
For $\epsilon_{v}/\epsilon_{h}>1$, the gas-dense-fluid (GDF) coexistence
appears and is independent of the density $\rho$. In the left panel,
the background shows a density plot of the entropy $\mathcal{S}/k_{B}$
at fixed temperature $k_{B}T/\epsilon_{h}=0.01$. In the BDF phase,
there is a maximum entropy given by $\mathcal{S}/k_{B}=\ln\left(\tfrac{\sqrt{5}+1}{2}\right)=0.481211825$,
which occurs at the density $\rho=\tfrac{1}{2}+\tfrac{\sqrt{5}}{10}\approx0.7236068$.
In contrast, there is no residual entropy in the GBF and GDF phases.
The right panel shows a density plot of the chemical potential $\mu/\epsilon_{h}$
in the same $\rho\text{\textendash}\epsilon_{v}/\epsilon_{h}$ plane
and at the same temperature. In the GDF and BDF phases, $\mu/\epsilon_{h}$
decreases with increasing $\epsilon_{v}/\epsilon_{h}$, while in the
GBF phase, it remains constant at $\mu/\epsilon_{h}=-1$.

This behavior highlights the subtle structure of phase coexistence
in this model. It is interesting to note that what may appear as an
interface between phases in GCE can manifest as a distinct phase in
canonical ensemble, and vice versa.

This behavior reveals the subtle nature of phase coexistence. Features
that appear as smooth transitions in the grand-canonical ensemble
may correspond to distinct thermodynamic states under fixed density,
and vice versa.

\subsection{Anomalous behavior under fixed density}

Let us now explore the anomalous behavior that emerges when the system
is analyzed at fixed particle density, following an approach similar
to that of Yasinskaya and Panov\,\citep{Yasinskaya}. In this constrained
setting, the pseudotransition manifests differently from its counterpart
in the GCE. Yasinskaya and Panov\,\citep{Yasinskaya} analyzed such
pseudotransitions in detail for a diluted spin chain by enforcing
density conservation via the Maxwell construction. In our case, we
access the fixed-density regime through an exact Legendre transformation
of the grand potential. This allows us to examine how the anomalies
in entropy and specific heat are modified when density is held constant,
highlighting the role of thermodynamic constraints in shaping the
observable features of the pseudotransition.

\begin{figure}
\includegraphics[scale=0.56]{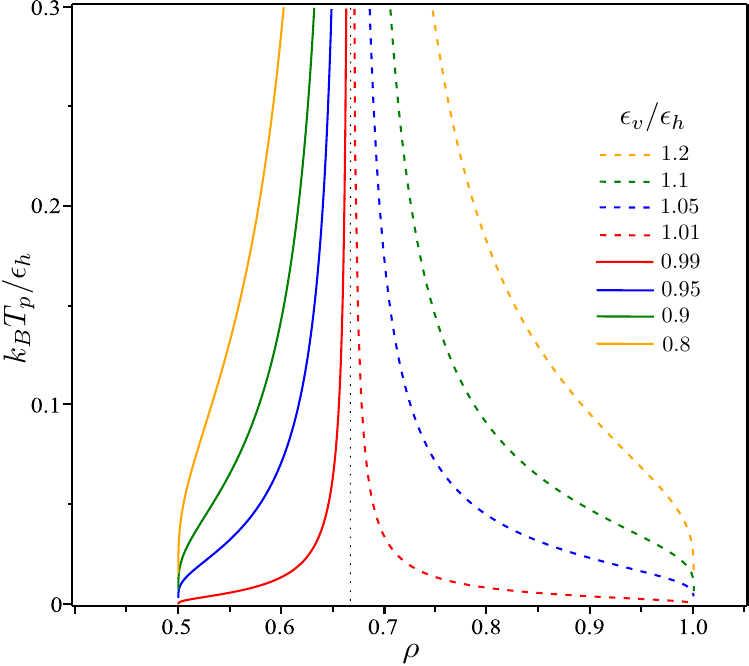}\caption{Pseudo-critical temperature $k_{B}T_{p}/\epsilon_{h}$ as a function
of molecular density $\rho$, calculated using Eq. \eqref{eq:Tp-rho}. }
\label{fig:Tp-rho} 
\end{figure}

Figure\, \ref{fig:Tp-rho} shows the pseudo-critical temperature,
given by Eq.\,\eqref{eq:Tp-rho}, as a function of the density $\rho$
for several fixed values of $\epsilon_{v}/\epsilon_{h}$, indicated
within the panel. The dimensionless temperature $k_{B}T_{p}/\epsilon_{h}$
increases rapidly as $\rho$ approaches $2/3$, marked by a thin vertical
dashed line. Solid lines represent the case $\epsilon_{v}/\epsilon_{h}<1$,
while dashed lines correspond to $\epsilon_{v}/\epsilon_{h}>1$. The
pseudo-critical temperature appears only within the density range
$0.5<\rho<1$. This provides an alternative perspective on the pseudotransition,
particularly around $\rho\approx2/3$. These results are equivalent
to those in Fig.\,\ref{fig:Pseudo-critical-T}, which displays $k_{B}T_{p}/\epsilon_{h}$
as a function of $\mu/\epsilon_{h}$.

\begin{figure}
\includegraphics[scale=0.62]{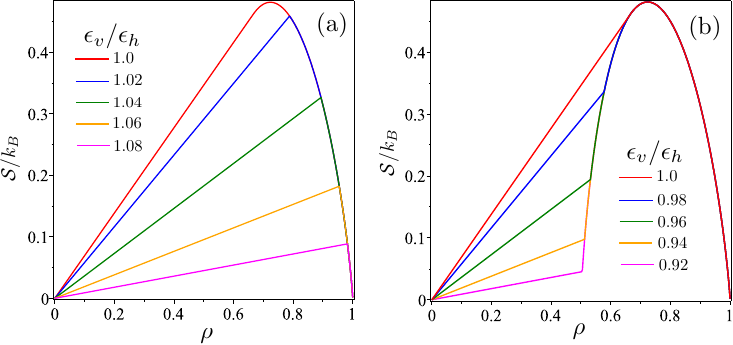}\caption{Entropy ${\cal S}/k_{B}$ as a function of density $\rho$, at fixed
temperature $k_{B}T_{p}/\epsilon_{h}=0.02$. (a) For $\epsilon_{v}/\epsilon_{h}>1$;
and (b) $\epsilon_{v}/\epsilon_{h}<1$}
\label{fig:Entropy-rho} 
\end{figure}

In Fig.\,\ref{fig:Entropy-rho}a, the entropy is shown as a function
of particle density $\rho$ for several fixed values of $\epsilon_{v}/\epsilon_{h}>1$,
indicated in the legend, at a constant temperature $k_{B}T_{p}/\epsilon_{h}=0.02$.
The red curve corresponds to the special case $\epsilon_{v}/\epsilon_{h}=1$.
The residual entropy can be obtained analytically by taking the limit
$\lim_{T\rightarrow0}\mathcal{S}(T,\rho)$, yielding a density-dependent
result: 
\begin{equation}
\mathcal{S}/k_{B}=\begin{cases}
\rho\ln(2); & 0\leqslant\rho<\frac{2}{3}\\
\ln\left[\frac{\left(2\rho-1\right)^{\left(1-2\rho\right)}\rho^{\rho}}{\left(1-\rho\right)^{\left(1-\rho\right)}}\right]; & \frac{2}{3}\leqslant\rho\leqslant1
\end{cases}.\label{eq:S_pcrit}
\end{equation}

The above expressions for the residual entropy can be derived using
a direct count of the number of accessible microstates. For low densities
and in the degenerate case of $\epsilon_{v}/\epsilon_{h}=1$, all
particles form a single aggregate. However, between each pair of particles,
one has or does not have a single empty site with the same probability.
Therefore, when placing each particle along the chain after a seed
particle is deposited at one of the chain borders, each new particle
can occupy two possible positions, namely, as a first-neighbor or
second-neighbor of the last-placed particle. At density $\rho$, the
total number of possible configurations is $\Omega=2^{\rho N}$, thus
resulting in the first expression in Eq.\eqref{eq:S_pcrit} that holds
for low densities. This form of distribution of articles results in
a cluster with an average size $3\rho N/2$ and, as such, can only
be built for $\rho<2/3$. A new configurational distribution of particles
develops for larger densities. Once the size of the cluster reaches
the chain size, between each pair of the ${\rho N}$ particles, one
can place or not one of the remaining $N-\rho N$ empty sites.The
number of possible configurations is now 
\begin{equation}
\Omega=\frac{(\rho N)!}{\left[(N-\rho N)![\rho N-(N-\rho N)\right)]!},
\end{equation}
which results in the second expression for the residual entropy in
Eq.\eqref{eq:S_pcrit}. Notice that the same residual entropy holds
for $\epsilon_{v}/\epsilon_{h}<1$ and $\rho>1/2$. The origin of
the characteristic densities and entropies reported in the preceding
section can now be seen. $\rho=2/3$ is the maximum density for which
the first of the two above configurations can be realized with the
maximum entropy per site being ${\mathcal{S}}/k_{B}=(2/3)\ln{(2)}$.
The maximum entropy within the second configuration is $\mathcal{S}/k_{B}=\ln\left(\tfrac{\sqrt{5}+1}{2}\right)$
occurring at $\rho=\tfrac{1}{2}+\tfrac{\sqrt{5}}{10}$.

In this formalism, the residual entropy is a function of $\rho$.
It is worth noting that the red curve in Fig.\,\ref{fig:Entropy-rho}a
does not clearly show a zero-temperature phase transition despite
being plotted at $T=0$, it is indistinguishable from the case at
$k_{B}T_{p}/\epsilon_{h}=0.02$. The remaining curves correspond to
the dashed lines in Fig.\,\ref{fig:Tp-rho}, and all are plotted
at $k_{B}T_{p}/\epsilon_{h}=0.02$. Each jagged (but continuous) curve
indicates a pseudotransition occurring for $\rho>2/3$. For increasing
values of $\epsilon_{v}/\epsilon_{h}>1$, the pseudo-critical density
shifts toward $\rho\rightarrow1^{-}$. Below this pseudo-critical
density $\rho_{p}$ (obtained from eq.\eqref{eq:Tp-rho}), the entropy
follows approximately the expression in \eqref{eq:S_pcrit}, while
above $\rho_{p}$, it transitions to the entropy associated with another
limiting form, also described in \eqref{eq:S_pcrit}.

Panel (b) of Fig.\,\ref{fig:Entropy-rho} presents similar entropy
density curves, but for $\epsilon_{v}/\epsilon_{h}<1$. In this case,
the pseudotransition appears within $0.5<\rho<2/3$. It is also worth
pointing out that the ground-state phases illustrated in Fig.\,\ref{fig:Phase-diagram}
can be identified in Fig.\,\ref{fig:Phase-diagram}a at the extreme
points: $\rho=0$ (G phase) and $\rho=1$ (DF phase). In panel (b),
the BF phase is visible at $\rho=0.5$.

\begin{figure}
\includegraphics[scale=0.65]{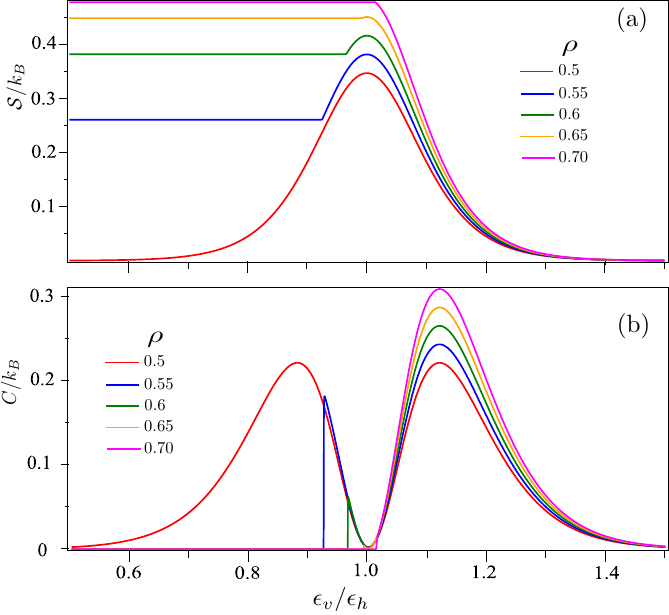}\caption{(a) Entropy as a function of $\epsilon_{v}/\epsilon_{h}$, at fixed
temperature $k_{B}T/\epsilon_{h}=0.05$. (b) Specific heat as a function
of $\epsilon_{v}/\epsilon_{h}$ }
\label{fig:S-C-ev} 
\end{figure}

Figure \ref{fig:S-C-ev}a shows the entropy as a function of $\epsilon_{v}/\epsilon_{h}$
for several fixed values of $\rho$, indicated in the panel. For $\rho=0.5$,
a pronounced and smooth peak appears at $\epsilon_{v}/\epsilon_{h}=1$
(red curve). For $0.5<\rho<2/3$, a kink emerges at the pseudo-critical
point: below this point, the entropy remains nearly constant, following
\eqref{eq:S_pcrit}, while above it, the entropy increases with $\epsilon_{v}/\epsilon_{h}$,
reaches a maximum, and then decreases. Figure \ref{fig:S-C-ev}b displays
the specific heat $C$ as a function of $\epsilon_{v}/\epsilon_{h}$
under the same conditions as panel (a). For $\rho=0.5$, the curve
exhibits two symmetric peaks around $\epsilon_{v}/\epsilon_{h}=1$,
characteristic of a standard zero-temperature phase transition. For
densities in the range $0.5<\rho<2/3$, the specific heat still reflects
the influence of the phase transition for $\epsilon_{v}/\epsilon_{h}<1$,
forming a jagged yet continuous structure. However, for $\rho>2/3$,
the influence of the zero-temperature phase transition fades on the
left side ($\epsilon_{v}/\epsilon_{h}<1$), with only the right side
($\epsilon_{v}/\epsilon_{h}>1$) retaining visible standard features
of the zero temperature phase transition.

\begin{figure}
\includegraphics[scale=0.6]{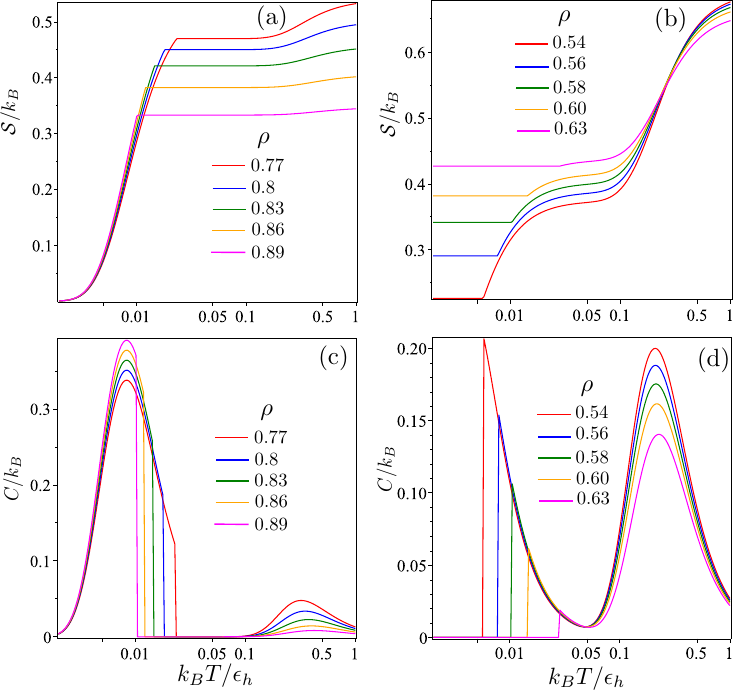}\caption{(a) Entropy as a function of temperature for fixed $\rho$ values
(indicated in the panel) with $\epsilon_{v}/\epsilon_{h}=1.02$. (b)
Same as (a), but for $\epsilon_{v}/\epsilon_{h}=0.99$. (c) Specific
heat versus temperature for the same $\rho$ values as in (a), assuming
$\epsilon_{v}/\epsilon_{h}=1.02$. (d) Same as (c), but with $\epsilon_{v}/\epsilon_{h}=0.99$.}
\label{fig:SC-rho-T} 
\end{figure}

Figure \ref{fig:SC-rho-T}a shows the entropy as a function of temperature
for fixed $\epsilon_{v}/\epsilon_{h}=1.02$ and several densities
in the interval $2/3<\rho<1$. We observe that the entropy starts
from zero, undergoes a kink at the pseudo-critical temperature $k_{B}T_{p}/\epsilon_{h}$,
and then increases smoothly, resembling the behavior reported for
the diluted Ising model \citep{Yasinskaya}. A similar trend is presented
in panel (b) for densities $0.5<\rho<2/3$. In this case, the entropy
begins with a residual value $\mathcal{S}=k_{B}\rho\ln(2)$, remains
nearly constant until it reaches $k_{B}T_{p}/\epsilon_{h}$, where
a kink occurs, followed by a standard increase. Panel (c) displays
the specific heat, defined as $C=\partial\mathcal{S}(T,\rho)/\partial T$,
under the same conditions as in panel (a). Here, the specific heat
exhibits a sharp drop at the pseudo-critical temperature, becoming
almost zero, and then increases as expected at higher temperatures
before eventually decaying asymptotically. Similarly, panel (d) reports
the specific heat under the conditions of panel (b). In this case,
the specific heat remains nearly zero up to the pseudo-critical temperature,
after which it rises sharply, reaches a minimum, increases again,
forming a smooth peak, and finally vanishes asymptotically.

\section{Conclusion}

This study examined a one-dimensional water-like model with van der
Waals and hydrogen bond interactions, incorporating particle number
fluctuations via a chemical potential. The model, introduced in \citep{Barbosa2013},
represents a simplified model for confined water on a linear chain
with periodic boundaries. At zero temperature, it presents three distinct
phases: gas, bonded liquid, and dense liquid; separated by clear boundaries
in the ($\mu,\,T$) plane. We extended the original analysis by focusing
on finite-temperature anomalies and on the influence of thermodynamic
constraints.

Using the transfer matrix method, we derived exact solutions in the
GCE and observed a rich thermodynamic structure marked by pseudotransitions.
Analytical expressions obtained from a cubic equation allowed us to
explore finite-temperature anomalies not previously addressed. These
anomalies, though analytic, closely resemble first- and second-order
phase transitions, with sharp changes in entropy, density, and internal
energy, along with finite peaks in specific heat and correlation length.
The latter's sharpness reinforces their interpretation as emergent
collective behavior without true criticality.

We also examined the thermodynamic behavior at fixed particle density
by applying a Legendre transformation to the grand-canonical solution.
Under this constraint, pseudotransition signatures become smoother.
Entropy shows a kink in curvature, and specific heat presents a finite
but non-divergent jump. Compared to the unconstrained case, the anomalies
are less abrupt, which illustrates how thermodynamic constraints can
modulate the visibility and character of pseudotransitions. Even so,
the anomalous region remains well defined, and the pseudotransition
temperature can still be consistently identified.

Although the model is not parametrized from first-principles water
potentials, it captures key aspects of quasi-one-dimensional molecular
systems with directional constraints. These include the competition
between bonding-type and steric interactions. These constraints are
relevant to recent studies of water confined in carbon nanotubes \citep{nano-carbon},
where similar quasi-critical thermodynamics behavior has been observed.
Our results suggest that such systems may exhibit pseudo-transitions
even in the absence of long-range order.

Overall, we found that the emergence and character of pseudotransitions
in one-dimensional systems depend not only on the microscopic Hamiltonian
but also on the thermodynamic conditions under which the system is
probed. Whether density is allowed to fluctuate or held fixed influences
how sectors of the state space contribute to the thermodynamics and
how collective anomalies manifest. Understanding this sensitivity
is essential for correctly interpreting pseudotransition phenomena
in low-dimensional and constrained systems. 
\begin{acknowledgments}
F.F.B. gratefully acknowledges the financial support from the Brazilian
agency CAPES for the project Assistive Technology for the Participation
of Persons with Disabilities in Exact Sciences (PDPG-AFIRMATIVA2600645P),
funded by the Graduate Program Development (PDPG) - Affirmative Policies
and Diversity, Call No. 17/2023, Thematic Alignment II. O.R. and S.M.S.
thank CNPq and FAPEMIG for partial financial support. M.L.L. also
acknowledges financial support provided by FAPEAL.
\end{acknowledgments}

\end{document}